\documentclass[
superscriptaddress,
 amsmath,amssymb,
 aps,
prl,
twocolumn,
]{revtex4-1}
\usepackage{graphicx}% Include figure files
\usepackage{dcolumn}% Align table columns on decimal point
\usepackage{bm}% bold math
\usepackage{natbib}
\usepackage{pgfplots}
\usepackage[utf8]{inputenc}  
\usepackage[T1]{fontenc}
\usepackage{dcolumn}     
\usepackage{hyperref}% add hypertext capabilities
\usepackage{ulem}
\begin{document}

\title{Surface tension and the strain-dependent topography of soft solids}

\author{Nicolas Bain}
\email{nicolas.bain@mat.ethz.ch}
 \affiliation{Department of Materials, ETH Z\"{u}rich, 8093 Z\"{u}rich, Switzerland.}
 \author{Anand Jagota}
 \email{anj6@lehigh.edu}
 \affiliation{Departments of Bioengineering, and of Chemical and Biomolecular Engineering, Lehigh University, Bethlehem, PA 18017, USA.}
 \author{Katrina Smith-Mannschott}
 \affiliation{Department of Materials, ETH Z\"{u}rich, 8093 Z\"{u}rich, Switzerland.}
 \author{Stefanie Heyden}
 \affiliation{Department of Materials, ETH Z\"{u}rich, 8093 Z\"{u}rich, Switzerland.}
 \author{Robert W. Style}
 \affiliation{Department of Materials, ETH Z\"{u}rich, 8093 Z\"{u}rich, Switzerland.}
 \author{Eric R. Dufresne}
\email{eric.dufresne@mat.ethz.ch}
 \affiliation{Department of Materials, ETH Z\"{u}rich, 8093 Z\"{u}rich, Switzerland.}
% Et dans les auteurs, mettre Camille Noûs, du laboratoire Cogitamus

\date{\today}

\begin{abstract}
When stretched in one direction, most solids shrink in the transverse directions.
In soft silicone gels, however, we observe that small-scale topographical features grow upon stretching.
A quantitative analysis of the topography shows that this counter-intuitive response is nearly linear, allowing us to tackle it through a small-strain analysis.
We find  that the surprising increase of small-scale topography with stretch is due to a delicate interplay of  the bulk and surface responses to strain. 
Specifically, we find that  surface tension changes as the material is deformed.
This response is expected on general grounds for solid materials, but challenges the standard description of gel- and elastomer-surfaces.
\end{abstract}

\pacs{Valid PACS appear here}
\maketitle

Surface tension is the driving force of a plethora of small-scale phenomena.
In liquids, it is responsible for the spherical shape of small drops and for the shape of  menisci \cite{de2013capillarity}.
In soft solids, surface tension rounds off sharp features \cite{lapinski2019surface,paretkar2014flattening,jagota2011adhesion,mora2013solid} and more generally governs mechanical responses at small scales \cite{style2017elastocapillarity,andreotti2020statics,mora2010capillarity,jerison2011deformation, style2013surface,style2013surface,style2013universal, park2014visualization,jensen2015wetting,barney2020cavitation}.
Despite the growing interest in soft solids for applications in microsystems and  robotics \cite{huh2010reconstituting,lacour2016materials,kim2013soft}, the essential nature of their surface properties remains elusive.

Generally, soft solids come in two forms, elastomers and gels.
Elastomers are made by lightly cross-linking a polymeric liquid.
Gels are cross-linked networks swollen with a liquid solvent.
In both cases, it is generally assumed that molecules can seamlessly rearrange at the surface, resulting in liquid-like surface properties. 
Specifically, it is expected that the surface tension is independent of the applied strain.
Recently, this assumption has been tested with wetting experiments.
Macroscopic experiments \cite{schulman2018surface} found no evidence for strain dependence of elastomers, but microscopic experiments \cite{xu2017direct,xu2018surface} found a marked increase of surface tension with applied strain, a characteristic feature of solid surfaces \cite{shuttleworth1950surface,gurtin1975continuum,haiss2001surface}.
While recent theoretical works have validated this experimental technique \cite{pandey2020singular,heyden2021contact},  others have called it into question because of the singular nature of the three-phase contact line \cite{masurel2019elastocapillary}.
Therefore, there is an urgent need for measurements of the strain-dependence of surface tension in soft solids that do not rely on wetting phenomena.

Here, we examine the strain-dependence of surface tension for a family of silicone solids, by quantifying their surface topography as a function of applied stretch.
We observe a counterintuitive increase of the amplitude of surface topography on soft silicone gels.
A small-strain analysis of this increase shows that it can only be quantitatively captured by  a  balance of strain-stiffening bulk properties and solid-like surface tension.

\begin{figure}[b]
\includegraphics[scale = 1]{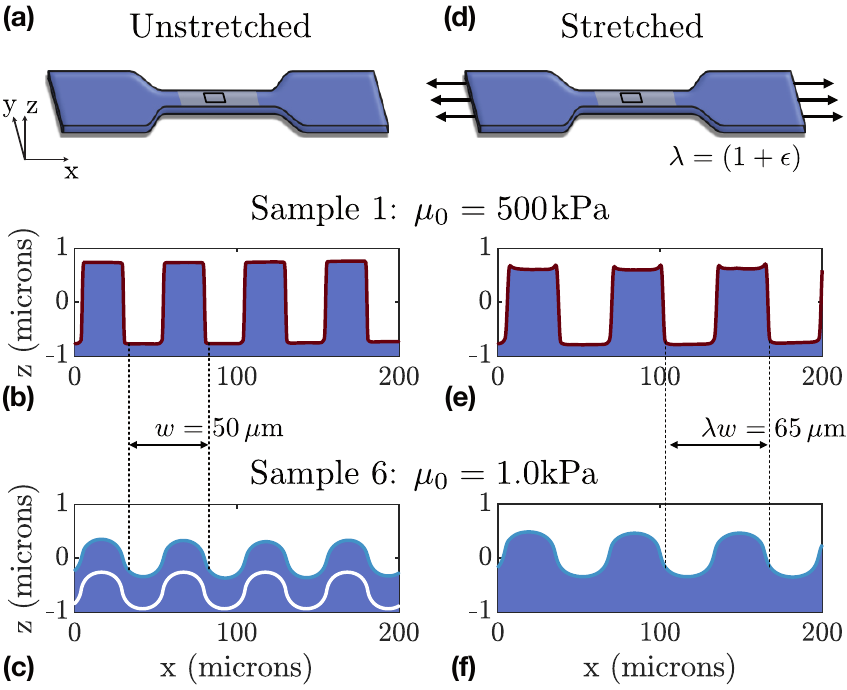}
\caption{\textit{The experimental setup.} (a) and (d) Schematic of the patterned dogbone samples. In (d), the stretch $\lambda$ and the strain $\epsilon$ are calculated from the change of the the pattern wavelength. (b) and (c) Averaged unstretched surface profile for the stiffest gel, Sample 1, and for the softest gel, Sample 6, respectively. The white line in (c) corresponds to the profile prediction from the linear flattening model, Eq.\eqref{eq_flattening}, applied to all Fourier modes of the initial profile (b), with shear modulus $\mu_0 = 1.0~\mathrm{kPa}$ and surface tension $\Upsilon_0 = 25.2~\mathrm{mN/m}$, Table~\ref{table_gels}. 
It agrees perfectly with the experimental profile and has been shifted down to be visible.
(e) and (f) Averaged surface profile for the stiffest and for the softest gel, respectively, stretched by 30\%. 
}
\label{fig_peak_to_peak}
\end{figure}

\begin{table*}[t]
\begin{ruledtabular}
\begin{tabular}{c|c|c|c|c}
 Name &Industrial Supplier& Shear modulus $\mu_0$&First mode $\tilde h_1$&Surface tension $\Upsilon_0$\\
 \hline
 Sample 1 & Sylgard184, Dow Corning Toray & $500\pm10~\mathrm{kPa}$ & $0.96 \pm .01~\mu\mathrm m$ & - \\
 Sample 2 & Sylgard184, Dow Corning Toray & $500 \pm 10~\mathrm{kPa}$ & $0.97 \pm .01~\mu\mathrm m$& - \\
 Sample 3 & DMS-V31/HMS-301 Gelest & $3.9 \pm 0.1 ~\mathrm{kPa}$ & $0.68 \pm .01 ~\mu\mathrm m$& $26.3 \pm 0.7~\mathrm{mN/m}$\\
 Sample 4 & CY52-276, Dow Corning Toray & $2.0 \pm 0.2 ~\mathrm{kPa}$ & $0.56 \pm .01~\mu\mathrm m$& $27.2 \pm 1.2~\mathrm{mN/m}$\\
 Sample 5 & CY52-276, Dow Corning Toray & $1.9 \pm 0.1~\mathrm{kPa}$ & $0.52 \pm .01 ~\mu\mathrm m$ & $21.8 \pm 1.1 ~\mathrm{mN/m}$\\
 Sample 6 & DMS-V31/HMS-301 Gelest & $1.0 \pm 0.1~\mathrm{kPa}$ & $0.38 \pm .01 ~\mu\mathrm m$ & $25.3 \pm 1.9~\mathrm{mN/m}$\\
\end{tabular}
\end{ruledtabular}
\caption{\textit{Summary table for the properties of all samples.} All mechanical properties are measured when unstretched. We measure the shear modulus $\mu_0$ from an independent indentation test, and the first Fourier mode $\tilde h_1$ from the surface topography. We measure the surface tension $\Upsilon_0$ from the linear flattening theory Eq.\eqref{eq_flattening}, using $h_0 = 0.97 \pm 0.01~\mu\mathrm m$ for reference, measured from the initial molds.
}
\label{table_gels}
\end{table*}

We cure polydimethylsiloxane (PDMS) gels inside dogbone-shaped poly(methyl methacrylate) molds, so that the central section of the dogbone has a $3\times 2$ mm$^2$ cross section and length of $20$ mm, Fig.~\ref{fig_peak_to_peak}a.
A periodic rectangular grating of wavelength $50$ $\mathrm{\mu m}$ and amplitude $1.52 \pm 0.01$ $\mathrm{\mu m}$ is applied to the surface during curing, to give the profile shown in Fig.~\ref{fig_peak_to_peak}b.
The grating is made of low-surface-tension chemically-inert fluoroplastic (3M DyneonTM Fluoroplastic Granules THV 500GZ) by melting fluoroplastic beads at 200$^\circ$C for 8 hours onto a stiff PDMS grating \cite{begolo2011new}.
After curing at 40$^\circ$C for a week to ensure full crosslinking, we detach the grating from the sample.
The patterned surface on the PDMS sample relaxes to a new shape, in which surface stresses and bulk elastic stresses balance \cite{hui2020surface,paretkar2014flattening,lapinski2019surface}, Fig.~\ref{fig_peak_to_peak}c.
We then stretch the dogbone-shaped sample, Fig.~\ref{fig_peak_to_peak}d, and measure the surface topography at each stretch state with a 3D optical profiler (S-neox, Sensoscan, $40
\times$ objective ).
The surface topography relaxes completely within a few minutes.
We image only the middle of the sample where we expect uniform stretching conditions.
We conduct the experiment on six silicone samples with different stiffnesses, each measured independently with an indentation test. 
For each sample, all measurements are at roughly the same location.
We name them ``Sample 1'' to ``Sample 6'' in order of decreasing stiffness (see Table \ref{table_gels} for properties and Supplement Section 1 \cite{supp} for experimental details). 

To develop a qualitative understanding of the results, we first  contrast the surface profiles of the stiffest and softest samples, Fig.~\ref{fig_peak_to_peak}. 
Surface profiles for the other gels are provided in Supplement Section 1.3 \cite{supp}. 
Each profile is an average of one topography measurement along the transverse direction $y$. 
Unstretched, the stiffer sample nicely reproduces the topography of the initial mold, Fig.~\ref{fig_peak_to_peak}b.
This is the behavior we expect from a stiff solid, which is stiff enough to resist significant deformation by surface tension.
Conversely, the topography of the softer gel strongly deviates from the one of the initial mold.
The peak-to-peak amplitude is halved, and the shape is  significantly rounded-off, Fig.~\ref{fig_peak_to_peak}c.

To explain this behavior, we begin with a force balance at the solid-air interface of an \textit{unstretched} solid.
Akin to the experiments, we consider a solid with an undulating surface, periodic in the $x$  direction and invariant along the $y$ direction, Fig.~\ref{fig_peak_to_peak}.
For simplicity we assume the initial surface profile to be sinusoidal, $h_0\cos qx$, where $q = 2 \pi / w$ is the pattern's wavevector and $h_0$ the surface amplitude when the sample is attached to the mold.
When released from the mold, the surface deforms into its new equilibrium profile $h_f \cos qx$.
We estimate the final amplitude $h_f$ by balancing the vertical stresses on both sides of the solid-air interface.
On the bulk side, the normal displacement of the surface $v = (h_f - h_0)\cos qx$ creates a normal stress response $\sigma_z$.
For an isotropic incompressible linear elastic solid, this stress is proportional to the shear modulus $\mu_0$: $\sigma_z = 2\mu_0|q|v$ \cite{johnson1987contact}.
On the other hand, surface tension $\Upsilon_0$ creates a jump in the stresses across the undulating interface, $\sigma_\Upsilon$, proportional to the local curvature of the final profile: $\sigma_\Upsilon = - q^2 \Upsilon_0 h_f\cos qx$ \cite{style2017elastocapillarity}.
The solid is at mechanical equilibrium when the jump in stresses caused by surface tension is equal to the stress response from the bulk deformation $\sigma_\Upsilon = \sigma_z$.
From this stress balance we express the final amplitude as a function of the initial one
\begin{equation}
	h_f = \frac{h_0}{1 + |q|\frac{\Upsilon_0}{2\mu_0}},
	\label{eq_flattening}
\end{equation}
which we call the flattening equation \cite{paretkar2014flattening,lapinski2019surface,hui2020surface}. 
Assuming linear response, this process can be applied to any Fourier mode of a non-sinusoidal surface.
Simply stated, surface tension acts as a low-pass filter
with a cut-off length equal to the elastocapillary length $\Upsilon_0 / 2\mu_0$.

We use this result to determine the surface tension of the soft gels (3-6) by inverting the flattening equation Eq.~\eqref{eq_flattening} for the first Fourier mode of the unstretched samples.
For all the soft gels we obtain nearly the same unstretched surface tension, $\Upsilon_0 = 25.2 \pm 2.4~\mathrm{mN/m}$, Table \ref{table_gels}. 
This value is close to $\gamma = 21\pm 1~\mathrm{mN/m}$, the surface tension of uncrosslinked PDMS \cite{bergeron1996monolayer}.
To further validate this model, we apply the flattening equation to each of the Fourier modes of the mold  for the softest sample, and nicely recover the experimental  profile, as shown by the light curve in Fig.~\ref{fig_peak_to_peak}c.

We now return  to the qualitative comparison of the softest and stiffest samples,  and consider the effect of stretching on their surface profiles, Fig.~\ref{fig_peak_to_peak}e and Fig.~\ref{fig_peak_to_peak}f.
With stretch, the wavelength of both surface profiles increases.
The peak-to-peak amplitude of the stiffest gel goes down, as expected due to its near-incompressibility, Fig.~\ref{fig_peak_to_peak}e.
Surprisingly, the peak-to-peak amplitude of the softest gel \textit{increases}, even though it is also nearly incompressible, Fig.~\ref{fig_peak_to_peak}f.

\begin{figure}[b]
\includegraphics[scale = 1]{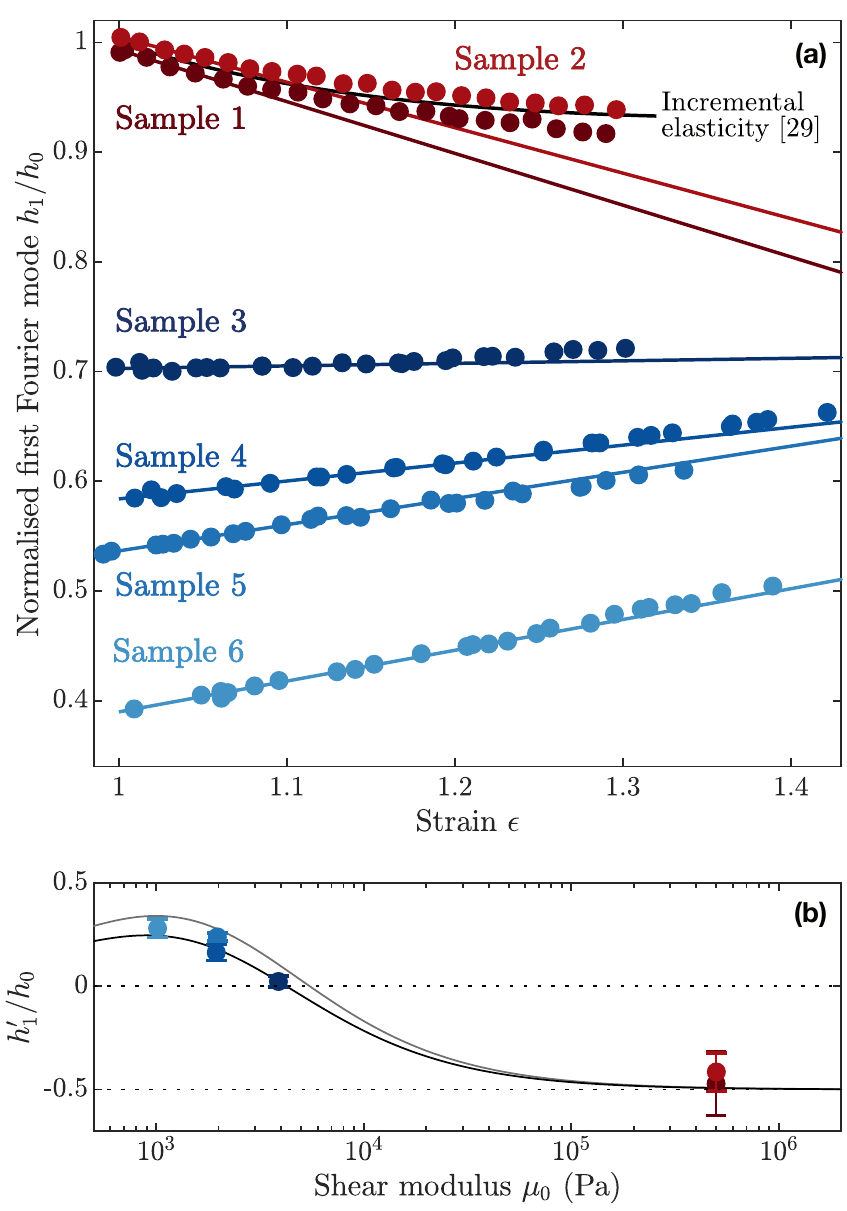}
\caption{\textit{Quantification of the strain-dependent topography} (a) Strain-dependent topography of the first Fourier mode amplitude for all samples as a function of strain $\epsilon$. Circles: experimental data. 
Errorbars are smaller than marker size.
Solid colored lines: linear fits of the data at small strains, for up to 15\% for samples (3-6) and up to 5\% for samples (1-2).
Black line: prediction from incremental elasticity (see Supplement Section 2 \cite{supp}).
(b) Initial slope of the strain-dependent first Fourier mode, as measured in (a), for all samples as a function of shear modulus $\mu_0$. Errorbars correspond to the 95\% confidence interval of the linear fit.
Black lines: prediction of the initial slope from linear expansion Eq.\eqref{eq_initial_slope} with surface elasticity $s=1$, bottom curve, and no surface elasticity $s=0$, top curve.
Blue line: prediction of the initial slope from linear expansion Eq.\eqref{eq_initial_slope} with surface elasticity $s=0.4$.
These predictions were made for $\alpha = 1/2$ and $b = 5/4$. 
}
\label{fig_first_mode}
\end{figure}

To quantify the strain-dependent topography of stiff and soft gels, we Fourier decompose the surface profile at each stretch state. 
We measure the strain $\epsilon$ from the difference between the wavelength of the periodic grating at each stretch state and the initial wavelength of $50$ microns, Fig.~\ref{fig_peak_to_peak}.
The amplitude of the first Fourier mode for all six gels, $\tilde h_1$, are shown in Fig.~\ref{fig_first_mode}a.
Each measurement is repeated six times to suppress  contributions from environmental noise.
While the amplitude of the first mode decreases monotonically with stretch for the MPa-scale elastomers, for all the kPa-scale gels the amplitude of the first mode exhibits the unexpected increase with applied stretch.

The response of the first Fourier mode amplitude to stretch is surprisingly linear, especially for the softer samples, Fig~\ref{fig_first_mode}a.
Therefore, we quantify the stretch response with the initial slope $h'_1 = \mathrm{d} h_1 / \mathrm{d} \epsilon$ from a linear fit at small strains, up to 15\% for the kPa samples and up to 5\% for the MPa samples, Fig~\ref{fig_first_mode}b.
The  slope decreases from a positive value for the softest samples to a negative value for the stiffest one, crossing zero at a shear modulus around $5~\mathrm{kPa}$.

To elucidate these counterintuitive observations, we investigate the competition of surface tension and bulk elasticity during stretching.
If we assume the flattening process to be linear, we can apply it as a perturbation to a stretched solid initially devoid of surface tension.
Within this assumption, stretching the solid accounts to changing  Eq.~\eqref{eq_flattening} in four ways.
First, the initial surface amplitude $h_0$ is replaced by its stretched counterpart $h_\epsilon$.
Then the period of the surface profile lengthens with strain, and the associated wavevector becomes $q / (1 + \epsilon)$.
If solid-like, the value of the surface tension will also be strain-dependent $\Upsilon_\epsilon(\epsilon)$. 
Finally, the bulk elastic properties are also subject to change, responding to a sinusoidal vertical displacement with a strain-dependent effective shear modulus $\mu_\epsilon(\epsilon)$.
In short, all the terms of the flattening equation Eq.\eqref{eq_flattening} become strain-dependent
\begin{equation}
	h_f(\epsilon)= \frac{h_\epsilon}{1 + \left|\frac{q}{(1 + \epsilon)}\right| \frac{\Upsilon_\epsilon}{2\mu_\epsilon}}.
	\label{eq_flattening_strain_dependent}
\end{equation}

We now conduct a linear expansion around small strains, $\epsilon \ll 1$.
At first order, the final amplitude is linear in strain $h_f(\epsilon) = h_f(0) + h'_f \epsilon + \mathcal O(\epsilon^2)$, with unstrained amplitude, $h_f(0)$, and initial slope, $h'_f$, measured in Fig~\ref{fig_first_mode}b.
Linearising each term of the strain-dependent flattening equation \eqref{eq_flattening_strain_dependent},  $h_\epsilon = h_0(1 - \alpha \epsilon) + \mathcal O(\epsilon^2)$, $\mu_\epsilon = \mu_0(1 + b \epsilon) + \mathcal O(\epsilon^2)$ and $\Upsilon_\epsilon = \Upsilon_0(1 + s \epsilon) + \mathcal O(\epsilon^2)$, we find their respective contributions to the initial slope
\begin{equation}
	h'_f = h_0\frac{\left(|q| \frac{\Upsilon_0}{2\mu_0}\left[1 - \alpha + b - s \right] - \alpha\right)}{\left(1 + |q|\frac{\Upsilon_0}{2\mu_0}\right)^2}. 
	\label{eq_initial_slope}
\end{equation}
In the limit where the bulk and surface properties are strain-independent  (\textit{i.e.} $s=b=0$), we see that positive initial slopes are only possible when surface tension is sufficient to overwhelm the impact of the bulk-term, $\alpha$.  The parameter $\alpha$ characterises the strain-dependent amplitude of a patterned solid in the absence of surface tension.
In the geometry of the present experiment, imposing a longitudinal strain to a slender beam, incompressibility requires $\alpha = 1/2$, which is in good agreement with the initial slope of the MPa samples Fig~\ref{fig_first_mode}b.
When the materials properties depend on the strain (\textit{i.e.} $s,b \ne 0$), contributions from the  surface and bulk have opposing effects: while the bulk strain-stiffening ratio $b$ increases the slope, surface elasticity ratio $s$ decreases it.

To isolate the effect of  surface elasticity, we  must first  evaluate the strain-dependent response of the bulk.
To do so, we first need to identify the appropriate nonlinear constitutive relation for our materials.  
As shown in the Supplement Section 1.2 \cite{supp}, our silicone based materials are well modelled as incompressible Neo-Hookean solids over the current range of strains.
In our experimental geometry, incremental elasticity  \cite{biot1965mechanics} gives a small-strain stiffening ratio of $b = 5/4$, as derived in Supplement Section 2 \cite{supp}.

\begin{figure}[!]
\includegraphics[scale = 1]{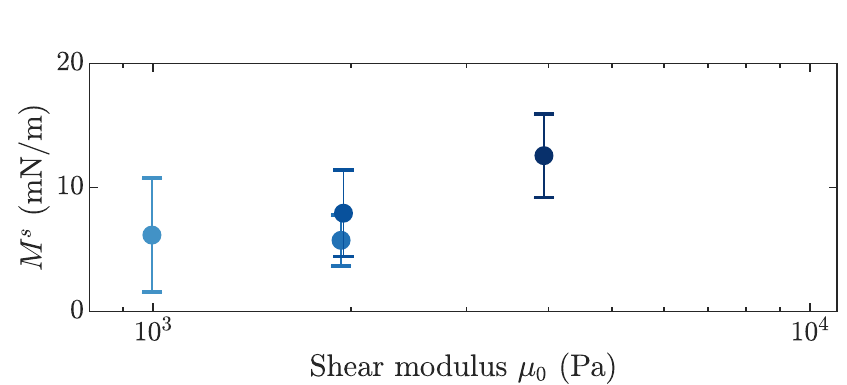}
\caption{\textit{Surface and bulk elastic constants.} Surface longitudinal modulus $M^s$ as a function of shear modulus $\mu_0$ for all four soft samples.
The error bars correspond to the 95\% confidence interval of the linear fits in Fig.~\ref{fig_first_mode}.
}
\label{fig_surf_el}
\end{figure}

With this result, we can now investigate the role of surface elasticity.
We plot predicted values of $h'_f/h_0$ versus shear modulus in 
Fig.~\ref{fig_first_mode}b.
While liquid-like surface tension (gray curve, $s=0$) systematically overestimates the initial slope, we find better agreement between theory and experiment for $s=0.4$.
Physically, we can interpret non-zero $s$ values as surface elasticity, with an elastic constant $M^s = s \Upsilon_0$.
We can independently determine the value of $M^s$  from the measured values of $h'_f$ in each experiment, using   Eq. \eqref{eq_initial_slope}.
The resulting values of $M^s$ are plotted for all the soft samples in Fig.~\ref{fig_surf_el}.
Similar to \cite{xu2017direct, xu2018surface}, we find a non-zero surface modulus for  all the soft gels.
However, we find a weaker effect of surface elasticity.  
In this experiment, we find $M^s \approx 10~\mathrm{mN/m}$.
According to the wetting measurements of \cite{xu2018surface,heyden2021contact}, we would expect $M^s=52~\mathrm{mN/m}$, see Suplement Section 4 \cite{supp}.
Interestingly, these data suggest that the surface elastic consant increases with the bulk network stiffness,  echoing numerical simulations of polymeric networks \cite{liang2018surface}.

We have found that the strain-dependent topography of soft silicone gels is consistent with non-zero surface elastic constants.
Qualitatively, we reach the same conclusion as previous wetting experiments \cite{xu2017direct,xu2018surface}.
Quantitatively, we find significantly smaller surface elastic moduli.
This discrepancy could arise from nonlinear stress focusing at sharp wetting ridges \cite{masurel2019elastocapillary,pandey2020singular}, or differences in surface preparation. 
\footnote{In the wetting experiments, silicone was cured in air. 
Here, silicone was cured against a teflon surface and removed by peeling}.
While both experiments couple surface elasticity and bulk nonlinearities, the singular behavior of a three-phase contact line complicates their decoupling \cite{masurel2019elastocapillary,pandey2020singular}.
The present experiment avoids this singularity, allowing us to develop a simple framework to disentangle bulk and surface effects.
More generally, our experiments suggest that the surface mechanical properties of swollen polymer networks can have significant contributions from both their solvent and network.  
They  raise basic questions in physics of polymer networks that have received limited theoretical attention.
How is the structure of the polymer network different at the surface than in the bulk?
What sets the magnitude of  surface elastic moduli? 
Can they be tuned independently of bulk elastic moduli?
Can we engineer surface elastic constants by curing in different environments, or through the addition of surface-active species, as done with liquid-liquid interfaces \cite{champougny2015surfactant,pepicelli2017characterization,fuller2012complex}?
We hope that the answers to these questions will provide fresh insights into the physics of soft solids, and enable future applications in wetting and adhesion.

\paragraph{Acknowledgements:} 

We acknowledge support from ETH Zürich Post-doctoral fellowship. We thank J. Snoeijer, J. Dervaux and T. Salez for insightful discussions. We also thank the Isa Lab for letting us use their white light profilometer, as well as O. Dudaryeva for introducing us to the use of fluoroplastic to pattern silicone surfaces.

% \newpage

\setcounter{equation}{0}
\setcounter{figure}{0}
\setcounter{table}{0}
\renewcommand{\thefigure}{S\arabic{figure}}
\renewcommand{\theequation}{S\arabic{equation}}
\renewcommand{\thetable}{S\arabic{table}}

\section{Supplement section 1: Experimental supplementary material}

\subsection{Sample preparation}

We used the following recipes for the different samples.
Stiff samples (Samples 1 and 2): the stiff samples were prepared by mixing Sylgard 184 (Dow Corning) base with its curing agent at a 10:1 weight ratio.
Gelest gels (Samples 3 and 6): these soft gels were prepared by mixing uncrosslinked PDMS chains (DMS-V31, Gelest) with a chemical crosslinker (HMS-301, Gelest) and catalyst (SIP6831.2, Gelest).
CY52-276 gels (Samples 4 and 5): these soft gels were prepared by mixing CY52-276 (Dow Corning) Part A and Part B at a 1:1 weight ratio.

All gels were mixed with a centrifuge mixer (Hauschild SpeedMixer DAC 150.1 FV-K) at 3500 rpm for one minute, degassed in a vacuum chamber until no bubbles remained, poured into a dogbone-shaped acrylic mold with a fluoroplastic pattern atop, and cured at 40$^\circ$C for a week to ensure full cross-linking. 

For each sample, we cured the excess material in the mixing container along with the dogbone-shaped sample and kept it for mechanical characterisation.

\subsection{Mechanical characterisation}

We characterised the gels stiffness by conducting an indentation test on the leftover gel in the mixing container, with a texture analyser (TA.XTPlus, Stable Micro Systems).
We used a $R = 1\,\mathrm{mm}$ radius cylindrical probe and a 500g load cell.
We obtain the shear modulus $\mu_0$ from contact mechanics \cite{johnson1987contact}
\begin{equation}
\mu_0 = \beta\frac{(1 -\nu)}{4 R}F'(d),  
\end{equation}
where $\nu$ is the Poisson ratio, which we consider equal to $1/2$, $F'(d)$ is the slope of the force-displacement curve and $\beta$ is a correction factor that accounts for the finite thickness $t$ of the sample \cite{lin2000detailed}
\begin{equation}
\beta = \left(1 + \left(\frac{.75}{(R/t) + (R/t)^3} +  \frac{2.8(1-2\nu)}{(R/t)}\right)^{-1}\right)^{-1}.  
\end{equation}

\begin{figure}[h]
\includegraphics[scale=1]{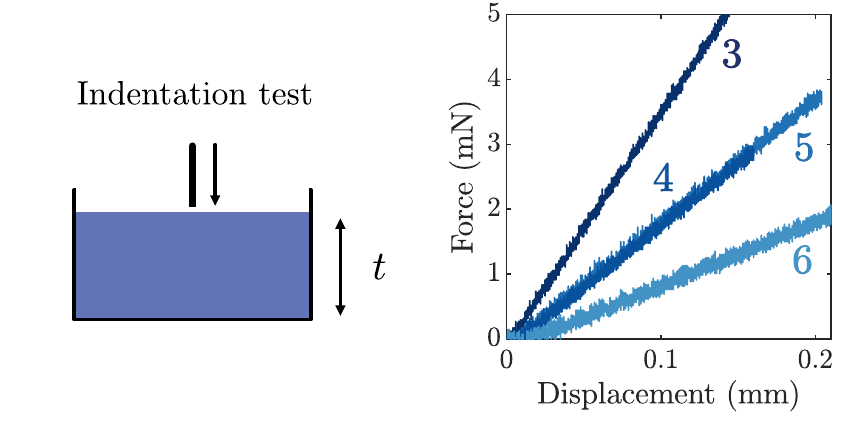}
\caption{Indentation test. Schematic of the indentation test, and force-displacement curve of an indentation test for samples 3 to 6. 
}
\label{fig_indentation}
\end{figure}

To remove environmental noise, we averaged the measurement of the slope $F'(d)$ over two repeated indentation tests at five sample locations.
We measured the sample thickness with callipers.
We plot in Fig.~\ref{fig_indentation} one force-displacement curve for each of the soft samples.

\begin{figure}[h]
\includegraphics[scale=1]{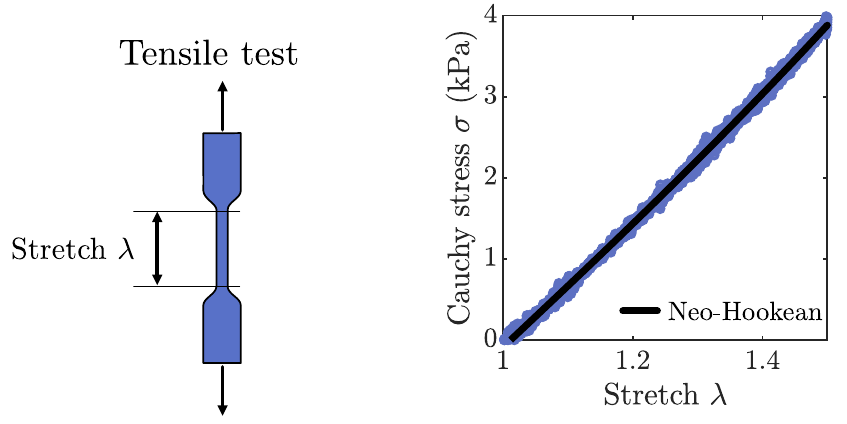}
\caption{Tensile test. Stress-stretch curve for a tensile test of a dogbone-shaped sample.
Solid line: theoretical behavior for an incompressible Neo-Hookean material.
}
\label{fig_mechanical_behavior}
\end{figure}

We also ensure that the soft gels behave as incompressible Neo-Hookean materials in the considered range of strains by conducting a tensile test.
We used a material testing machine (Zwick/Roell II) with extensometer (videoXtens) and a $1\,\mathrm{kg}$ load cell to test a dogbone-shape sample prepared exactly like Samples 4 and 5, only without surface pattern.
The test was conducted at a rate of $1\,\mathrm{cm}$ per minute, which we verified was slow enough not to show any dynamic effect.
Under uniaxial stretch conditions, we expect the Cauchy stress $\sigma$ of an incompressible Neo-Hookean solid to follow
\begin{equation}
\sigma = \frac{F}{A_0 \lambda^{-1}} = \mu_0\left(\lambda^2  - \frac 1 \lambda\right),  
\end{equation}
where $F$ is the measured force, $A_0$ the initial cross-sectional area and $\lambda$ is the stretch.
We plot in Fig.~\ref{fig_mechanical_behavior} the experimental stress-stretch curve, together with a fit of the expected behavior for $A_0 = 6\,\mathrm{mm^2}$ and shear modulus $\mu_0 = 2.5\,\mathrm{kPa}$.
The good agreement between the data and the theoretical curve shows that the silicone gels we used behave as incompressible Neo-Hookean materials in the considered range of strains.

% \newpage

\subsection{Stretched profiles}

\begin{figure}[h]
\includegraphics[scale=1]{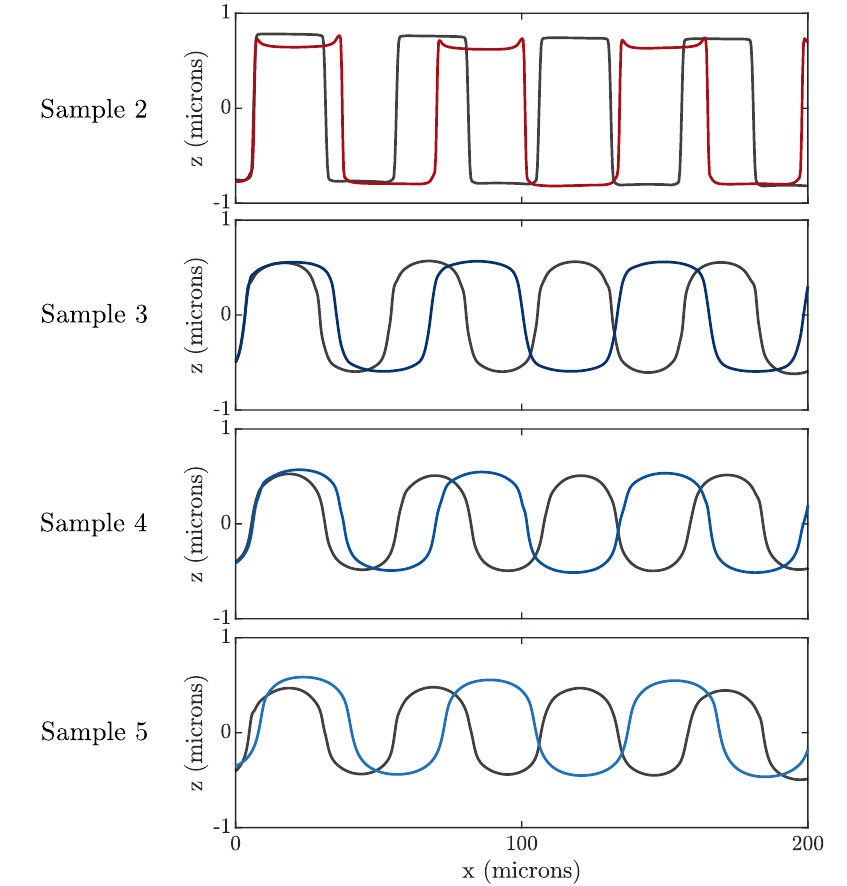}
\caption{Unstretched (black) and 30 \% stretched (colored) profiles for samples 2 to 5.}
\label{fig_all_profiles}
\end{figure}

We plot in Fig.~\ref{fig_all_profiles} the unstretched and stretched profiles for samples 2 to 5. The amplitude of the stiff sample decreases with stretch while the amplitude of all the soft ones increases with stretch. Samples 1 and 6 are shown in the main text.

% \newpage

\section{Supplement section 2: Incremental elasticity theory for strain-dependent topography}

Let us summarize the main steps of the incremental elasticity theory employed to establish the strain-dependent topography of stiff gels, Fig.~2a, and the strain-stiffening ratio $b$.

We consider a slab with a sinusoidal pattern on its top surface, identical along the $y$ direction (see Fig.~1 in the main text).
We first apply a uniaxial uniform pre-stretch and then consider how this state is perturbed by the presence of the surface topography.
We assume the slab to be made of an incompressible Neo-Hookean material.

\subsection{Uniaxial pre-stretch}

We apply a uniaxial stretch $\lambda_x = \lambda$.
From incompressibility, the two orthogonal stretches must satisfy $\lambda_y \lambda_z = 1 / \lambda$.
For an incompressible Neo-Hookean solid the pre-stretch principal Cauchy stresses obey $(S_i - S_j) = \mu_0(\lambda_i^2 - \lambda_j^2)$, where the indices $i$ and $j$ can be substituted by the principal stretch directions $x$, $y$ and $z$.

In the experimental conditions presented above, the uniaxial stretch of the dogbone guarantees $\lambda_y = \lambda_z = 1 / \sqrt \lambda$,
as well as no transverse stresses $S_y = S_z = 0$.
We can then express the longitudinal stress as a function of longitudinal stretch $S_x = \mu_0(\lambda^2 - 1 / \lambda)$.

The surface topography perturbs this uniform solution in two ways.
First, the undulating surface create local stress concentrations that will prevent the surface topography from following the uniform vertical stretch $\lambda_z$ \cite{gao1991stress}.
Second, if the solid is soft enough, surface tension will induce a
displacement of the surface in the $z$ direction, to which the solid will respond.
Before we study these two effects, we recall the main equations for perturbations to a pre-stretched state.

\subsection{Small perturbation to a pre-stretched solid}

We restrict ourselves to a 2D plane-strain perturbation of a pre-stretched solid.
In other words, taking a section in the $(x, z)$ plane, we assume this perturbation creates no additional deformation in the $y$ direction.
This is justified by the fact that perturbations at the free surface decay over a characteristic length of tens of microns. 
The system can therefore be considered as infinitely thick when it comes to deformations induced by the surface topography.
In addition we assume no vertical pre-stress $S_z = 0$.

We consider a small perturbation to the pre-stretched state, in the form of a displacement field $(u_x, u_z)$.
We regard the pre-stretch state as the reference configuration and the final state as the deformed configuration.
Under the assumptions just described, \cite{biot1965mechanics} establishes the constitutive equations
\begin{eqnarray}
  \label{eq_s11}
  s_{11} &=& 2 \mu_s e_{xx} + p\\
  \label{eq_s22}
  s_{22} &=& 2 \mu_s e_{yy} + p\\
  \label{eq_s12}
  s_{12} &=& 2 \mu_s e_{xy},
\end{eqnarray}
where $e_{ij} = (\partial_i u_j + \partial_j u_i) / 2$ are the incremental strains, $p$ the incremental pressure and $\mu_s = \mu_0 (\lambda^2 +\lambda_z^2) / 2$ the incremental shear modulus for an incompressible Neo-Hookean solid.
Here, the incremental stresses $\mathbf s$ are expressed in the deformed configuration, and the $(1,2)$ directions correspond to the $(x, z)$ directions rotated by $\omega = (\partial_x u_z - \partial_z u_x)/2$.

The mechanical equilibrium conditions for these incremental stresses are
\begin{eqnarray}
  \label{eq_stat_eq_1}
  \partial_x s_{11} + \partial_z s_{12} + S_x \partial_y \omega = 0\\
  \label{eq_stat_eq_2}
  \partial_x s_{12} + \partial_z s_{22} + S_x \partial_x \omega = 0
\end{eqnarray}
and the stresses exerted on a surface of normal $\mathbf e_z$ are
\begin{eqnarray}
  \label{eq_sigma_zz}
    \sigma_{zz} &=& s_{22}\\
  \label{eq_sigma_xz}
    \sigma_{xz} &=& s_{12} - S_xe_{xy}
\end{eqnarray}
in the pre-stretched configuration.

\subsection{Oscillating solutions}

We look for solutions of static equilibrium Eqs.\eqref{eq_stat_eq_1} and \eqref{eq_stat_eq_2} that are oscillating in the $x$ direction and vanishing at $z\to-\infty$.
With the constitutive law Eqs.\eqref{eq_s11}, \eqref{eq_s22} and \eqref{eq_s12} given above, such solutions are
\begin{eqnarray}
  \phi &=& q^{-2}\left(C_1 e^{q z} + C_2 e^{kq z}\right)\sin qx,\\
  p &=& - C_2 S_x k e^{kqz}\cos qx,
\end{eqnarray}
where $\phi$ is the stream function defined by $\partial_x\phi = u_z$ and $-\partial_z\phi = u_x$, and $k = \lambda / \lambda_z$.
$(C_1, C_2)$ are constants to be determined from boundary conditions.

\subsection{Topography-induced stress concentration}

We now investigate the first perturbation to the pre-stretched state: stress concentrations due to geometrical non-linearities.
Coming back to the unstretched surface profile $h_0\cos qx$ in the main text, we assume a two-step deformation process.
First, we deform the solid with uniform stretches $\lambda$ in the $x$ direction.
As we saw above, this leads to uniform stretch $\lambda_z$ in the $z$ direction.
We thereby obtain an intermediate profile $h_i\cos q_ix$ with amplitude $h_i = \lambda_z h_0$ and wavevector $q_i = q / \lambda$, as well as a uniform longitudinal stress $S_x = \mu_0(\lambda^2 - \lambda_z^2)$ \cite{biot1965mechanics}.
In this situation, however, the solid is not at rest.
With an oscillating normal vector, $\mathbf n(x) = \mathbf e_z - h'(x)\mathbf e_x$ at first order, the longitudinal stress $S_x$ creates a shear traction $t_{xz} = S_x q_i h_i \sin q_i x$ at the surface.
These are the stress concentrations described in \cite{gao1991stress}. 
They will make the surface relax to $h_\epsilon\cos q_ix$, which we want to determine.
Using $\sigma_{xz} = -t_{xz}$ as a boundary condition in Eq.\eqref{eq_sigma_xz} to bring the solid to equilibrium, together with an absence of vertical stress $\sigma_{zz} = 0$ in Eq.\eqref{eq_sigma_zz} we find the two constants to be
\begin{eqnarray}
  C_1 &=& -\left(\frac{2 \lambda \lambda_z(\lambda + \lambda_z)}{\lambda^3 + \lambda^2 \lambda_z + 3 \lambda \lambda_z^2 - \lambda_z^3}\right)q_i h_i\\
  C_2 &=& \left(\frac{(\lambda + \lambda_z)(\lambda^2 + \lambda_z^2)}{\lambda^3 + \lambda^2 \lambda_z + 3 \lambda \lambda_z^2 - \lambda_z^3}\right)q_i h_i,
\end{eqnarray}
giving a vertical sinusoidal displacement $u_z = V\cos q_i x$ of amplitude 
\begin{equation}
  V = \left(\frac{(\lambda + \lambda_z)(\lambda - \lambda_z)^2}{\lambda^3 + \lambda^2 \lambda_z + 3 \lambda \lambda_z^2 - \lambda_z^3}\right) h_i.
\end{equation}
Eventually, the amplitude at mechanical equilibrium is 
\begin{equation}
  h_\epsilon = h_i(1 + V) = h_i\left(\frac{2\lambda(\lambda^2 + \lambda_z^2)}{\lambda^3 + \lambda^2 \lambda_z + 3 \lambda \lambda_z^2 - \lambda_z^3}\right).
\end{equation}
Using the experimental conditions $\lambda_z = 1 / \sqrt \lambda$ and $h_i = h_0 / \sqrt \lambda$ for the uniform pre-stretch, we obtain
\begin{equation}
  h_\epsilon = h_0 \left(\frac{2(\lambda^4 + \lambda)}{\lambda^{9/2} + \lambda^3 + 3 \lambda^{3/2} - 1}\right).
  \label{eq_hs}
\end{equation}
At first order in strains $\epsilon = \lambda - 1$ we recover $h_\epsilon = h_0(1 - \epsilon / 2) +\mathcal O(\epsilon^2)$, as assumed in the main text.

\subsection{Effect of surface tension}

In addition to the vertical displacement induced by geometrical non-linearities, surface tension will also change the topography amplitude.
Starting now from a sinusoidal surface $h_\epsilon \cos q_i x$ at mechanical equilibrium, we assume surface tension will deform it into $h_f \cos q_i x$.
We use the incremental elasticity presented above to determine the stress response due to this vertical displacement $v = (h_f - h_\epsilon)\cos q_i x$.
Using $\partial_x\phi = v$ as a surface boundary condition and the absence of shear surface stress $\sigma_{xz} = 0$ as the other boundary condition in Eq.\eqref{eq_sigma_xz}, we obtain the two constants
\begin{eqnarray}
  C_1 &=& \left(\frac{\lambda^2 + \lambda_z^2}{\lambda^2 - \lambda_z^2}\right)q_i (h_f - h_\epsilon)\\
  C_2 &=& \left(\frac{2 \lambda_z^2}{\lambda_z^2 - \lambda^2}\right)q_i (h_f - h_\epsilon)
\end{eqnarray}
and a surface vertical stress
\begin{equation}
  \sigma_{zz} = \left(\frac{\lambda^3 + \lambda^2 \lambda_z + 3 \lambda \lambda_z^2 - \lambda_z^3}{\lambda + \lambda_z}\right)\mu_0|q_i| v.
  \label{eq_inc_s_zz}
\end{equation}
Comparing this response to the linear elastic response of $2\mu_0|q_i|v$, we define the effective shear modulus as
\begin{equation}
  \mu_\epsilon = \frac{\mu_0}2\left(\frac{\lambda^3 + \lambda^2 \lambda_z + 3 \lambda \lambda_z^2 - \lambda_z^3}{\lambda + \lambda_z}\right).
\end{equation}
On the other hand, the stress-jump created by surface tension is still equal to the product of surface tension and local curvature $\sigma_\Upsilon = -q_i^2 \Upsilon_{\lambda}h_f \cos q_i x$.
Equating it to the vertical stress due to the surface displacement Eq.\eqref{eq_inc_s_zz}, we recover the flattening equation with effective shear modulus $\mu_\epsilon$.
In the conditions of the experiments $\lambda_z = 1 / \sqrt \lambda$ we obtain the effective shear modulus
\begin{equation}
  \label{eq_mu_lambda}
  \mu_\epsilon = \frac{\mu_0}2 \left(\frac{\lambda^{9/2} + \lambda^3 + 3 \lambda^{3/2} - 1}{\lambda^{5/2} + \lambda}\right)
\end{equation}
which, at first order in strains $\epsilon = \lambda - 1$, gives a linearised effective modulus $\mu_\epsilon = \mu_0(1 + 5 \epsilon / 4)+\mathcal O(\epsilon^2)$.
We thereby obtain the strain-stiffening ratio $b = 5/4$ used in the main text.
We note that the effective shear modulus $\mu_\epsilon$ derived here is different from the strain-stiffened shear modulus $\mu_s$.
This is because the equilibrium equations solved in incremental elasticity differ from the ones of linear elasticity, resulting in an additional correction specific to each boundary-value problem \cite{biot1965mechanics}.
The value of the strain-stiffening ratio will therefore be problem dependent.

\section{Supplement section 3: Numerical validation}

\subsection{Numerical setup}

To validate the theory developed above, which couples a linear flattening process to incremental elasticity, we compare it with finite-elements simulations.
We create a quasi-2D geometry of length and height equal to $250$ microns and thickness of $12.5$ microns, with a surface rectangular pattern of wavelength $50$ microns and amplitude $1.5$ microns.
Akin to the experiment, we denote $(x,y,z)$ the directions corresponding to the width, thickness and height respectively (see Fig.~1 in the main text).
We apply symmetric displacements in the $x$ direction on both boundaries to stretch the system by $\lambda$.
We simultaneously apply a $1 / \sqrt \lambda$ stretch in the $y$ direction, and fix the bottom surface in the $z$ direction to avoid rigid-body motions.

\begin{figure}[h]
\includegraphics[scale=1]{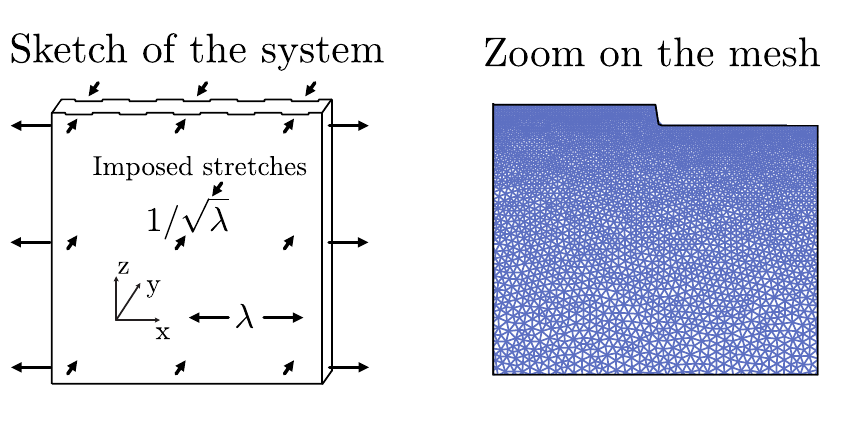}
\caption{Geometry of the numerical simulations.}
\label{fig_sketch}
\end{figure}

We solve the displacement fields with the a commercial finite-elements software (ABAQUS).
We define the material as slightly compressible Neo-Hookean, with Poisson ratio $\nu = 0.48$.
This slight compressibility is required for the numerics to converge.
We mesh the system with prismatic elements, triangular on each of the faces and with a depth equal to the system thickness.
We sketch in Fig.~\ref{fig_sketch} the geometry of the numerical simulations.

\subsection{Liquid-like surface tension}

We here consider the simplest surface state:
we assume surface tension to be liquid-like, strain-independent.
With this approach we can directly compare the numerical results to theoretical predictions.

\begin{figure}[h]
\includegraphics[scale=1]{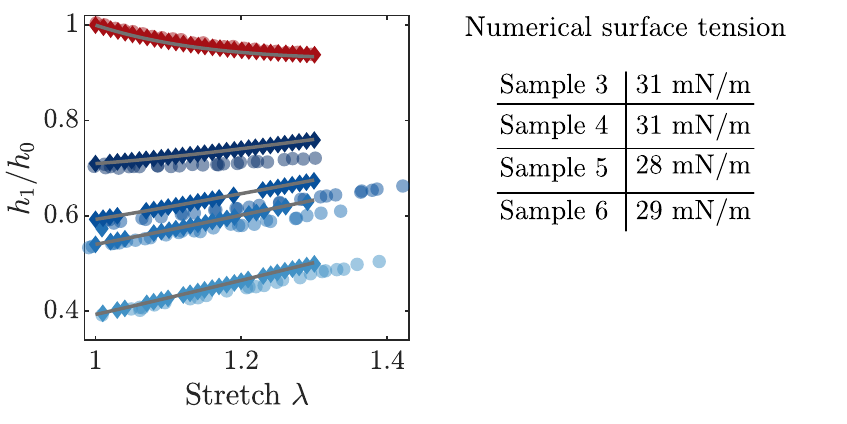}
\caption{Numerical results and theoretical predictions for liquid-like surface tension. Circles: experimental data. Diamonds: numerical simulations. We removed the simulations that did not converge. Solid lines: theoretical predictions.}
\label{fig_liquid_like}
\end{figure}

We apply liquid-like surface tension as a surface stress, constant in the deformed configuration, to the surface elements following \cite{lapinski2019surface}.
Application of surfaces stresses is implemented in ABAQUS via its User Element feature.  Specifically, this involves writing a subroutine that computes nodal forces and their derivatives and provides this to the main ABAQUS program.
We conduct five simulations in this scenario.
One with negligible surface tension, corresponding to our stiff samples, and one corresponding to each of our soft gels. 
We equate the numerical shear modulus to the experimental values, and choose the numerical surface tension value so that numerics and experiments have the same unstretched first mode amplitude.
We show in Fig.~\ref{fig_liquid_like} the prescribed values of numerical surface tension.
They are slightly higher than the experimental measurements (see main text, Table~1), a difference we attribute to the slight compressibility of the numerical model.

We recall that the theoretical prediction for the stiff sample is $h_\epsilon$, given in Eq.\eqref{eq_hs}, and that the prediction for soft samples are derived from $h_f(\epsilon)$, given in main text Eq.(2) in which $h_\epsilon$ and $\mu_\epsilon$ are described in Eqs.\eqref{eq_hs} and \eqref{eq_mu_lambda} respectively.

We plot in Fig.~\ref{fig_liquid_like} the experimental and numerical first mode amplitudes, as well as the theoretical predictions, for one stiff sample and the four soft samples.
The numerical results and the theoretical prediction are in perfect agreement over the whole range of stretch, and both show discrepancies with the experimental data: a liquid-like description of the surface overestimates the increase of the first Fourier mode with stretch.

This validates the theoretical model developed above, and confirms the results of the main text obtained in the small deformation regime: the strain-dependent topography of the studied silicone gels cannot be described with a strain-independent surface tension.

\section{Supplement section 4: Solid-like surface tension}

\subsection{Constitutive equation}

If a surface is solid-like, we need a constitutive equation to describe its mechanical strain response.
In the linear regime, such a law is described in \cite{xu2018surface},
\begin{equation}
  \Upsilon_{ij} = \Upsilon_0 + 2 \mu^s \epsilon^s_{ij} + \lambda^s \delta_{ij} \epsilon^s_{kk},
\end{equation}
where the surface stresses $\Upsilon_{ij}$ respond to surface strains $\epsilon^s_{ij}$ with two surface elastic constants, $\lambda^s$ and $\mu^s$.
As previously, $\Upsilon_0$ is the surface tension of an unstretched surface.
We note that the surface stresses are only isotropic if the surface strains are.

In the geometry of our experiment and in the linear regime, the longitudinal and transverse strains are related $\epsilon^s_{yy} = - \epsilon^s_{xx} / 2$.
We therefore obtain a constitutive equation
\begin{equation}
\label{eq_Upsilon_xx}
  \Upsilon_{xx} = \Upsilon_\epsilon = \Upsilon_0 + M^s \epsilon^s_{xx},
\end{equation}
where $M^s = (2 \mu^s + \lambda^s / 2)$ is the elastic constant measured in the main text.
In the wetting measurements, \cite{xu2018surface,heyden2021contact} measured $\mu^s = 12\,\mathrm{mN/m}$ and $\lambda^s = 55\,\mathrm{mN/m}$, which would give $M^s = 52\,\mathrm{mN/m}$ in the present geometry.

\subsection{Numerical simulations}

\begin{figure}[h]
\includegraphics[scale=1]{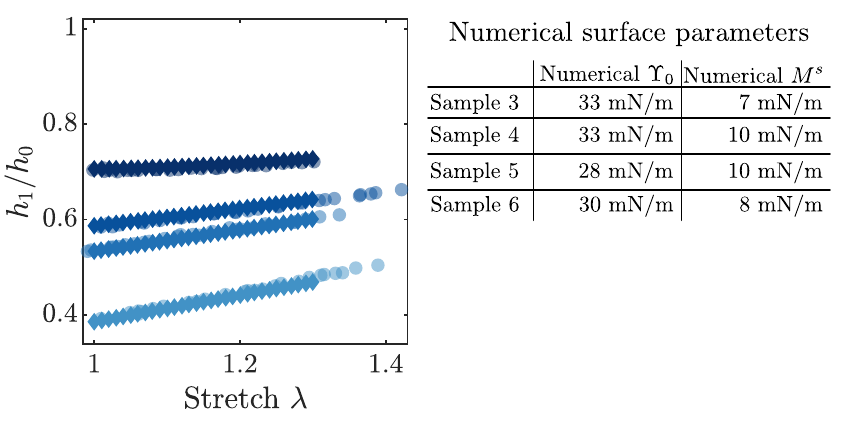}
\caption{Numerical results for solid-like surface tension. Circles: experimental data. Diamonds: numerical simulations.}
\label{fig_solid_like}
\end{figure}

To further validate our results we conduct numerical simulations as described above, but with a strain-dependent surface tension.
The surface constitutive law implemented in the finite element method has been formulated for isotropic behavior \cite{lapinski2019surface,milner1989buckling}.
The free energy $A$ is given by
 \begin{equation}
A=\Upsilon_0(J_s-1)+  \frac{M^s}{2}(J_s-1)^2      
\end{equation}
where $J_s$ the ratio of deformed to undeformed area.
For small strains, the corresponding surface stress magnitude reduces to Eq.\eqref{eq_Upsilon_xx} \cite{lapinski2019surface}.
We choose the numerical unstretched surface tension and the numerical surface modulus for the numerics to agree with the data for the four soft samples.
In Fig.~\ref{fig_solid_like} we show the agreement between numerics and data, as well as the chosen values for the numerical unstretched surface tension and surface elastic modulus.
All values are in good agreement with the experimental values measured in the small-strain regime.

\renewcommand\refname{Bibliography}
\bibliographystyle{apsrev4-1} 

\bibliography{paper_1_biblio}

\end{document}